\documentclass[a4paper,11pt]{article}
\usepackage[dvips]{graphicx}
\usepackage{amssymb}
\usepackage{amsmath}

\usepackage{cite}

\input arrow

\begin{document}

\title{Central Charge Extended Supersymmetric Structures for Fundamental Fermions Around non-Abelian Vortices}
\author{
K. Kleidis{\,} and V.K. Oikonomou\thanks{voiko@physics.auth.gr}\\
Department of Mechanical Engineering\\ Technological 
Education Institute of Serres \\
62124 Serres, Greece \\
} \maketitle

\begin{abstract}
Fermionic zero modes around non-abelian vortices are shown that they constitute two $N=2$, $d=1$ supersymmetric quantum mechanics algebras. These two algebras can be combined under certain circumstances to form a central charge extended $N=4$ supersymmetric quantum algebra. We thoroughly discuss the implications of the existence of supersymmetric quantum mechanics algebras, in the quantum Hilbert space of the fermionic zero modes.
\end{abstract}

\section*{Introduction}

Supersymmetry has played a prominent role in quantum field theory and sting theory model building the last thirty years. Regardless the fact that up to date no supersymmetric partner has been detected experimentally, the theoretical attributes of this graded super-Poincare algebra are so useful that made supersymmetry necessary in various fields of research. Since supersymmetry is not experimentally verified, this poses a compelling constraint that supersymmetry must be broken to our world. There exist various elegant ways to break supersymmetry in various contexts. For an important stream of papers regarding supersymmetry breaking in field theoretic grand unified theories see \cite{susy1,odi1,odi2}. In addition, for some cosmological and supergravity theories applications see \cite{odi3}. 

Supersymmetric quantum mechanics (abbreviated to SUSY QM hereafter), was introduced by Witten in order to find a simple description of supersymmetry breaking in quantum field theory \cite{witten1}. In time, SUSY QM has proved to be a powerful tool for studying dimensionally reduced quantum field theories, integrability in quantum mechanical systems and also was developed to be an independent research field. For important reviews and textbooks on SUSY QM, see \cite{reviewsusyqm} and references therein. SUSY QM is considered nowadays a research field which is by itself interesting and not only a convenient one dimensional theoretical tool for simplifying higher dimensional quantum field theories. The applications of SUSY QM are numerous and these applications cover a wide range of research areas. For example, mathematical aspects of Hilbert spaces corresponding to SUSY QM systems and also applications to various quantum mechanical systems were done in \cite{diffgeomsusyduyalities} and \cite{various,susyqminquantumsystems,plu1,plu2} respectively. Extended supersymmetries and harmonic superspaces were studied in \cite{extendedsusy,ivanov}, while applications of SUSY QM to scattering were presented in \cite{susyqmscatter}. Interesting features of supersymmetry breaking were studied in \cite{susybreaking}. The relation between central charge extended SUSY QM and global four dimensional spacetime supersymmetry was studied in \cite{ivanov}. In principle however, SUSY QM and four dimensional global supersymmetry are in general conceptually different.

In this paper, we focus on a particular interesting attribute of fermions around non-abelian vortices. Specifically, the zero modes of the fermions in the fundamental representation (isospinor fermions) constitute two independent $N=2$, $d=1$ supersymmetries, as we shall demonstrate. Under certain conditions, these supersymmetries can be combined to give an enhanced extended supersymmetric structure. Particularly they are combined to form an $N=4$, $d=1$ SUSY QM algebra with central charge. The origin of zero modes around vortices and solitonic backgrounds is a long standing riddle, as pointed out in \cite{rossi}. On the contrary, the existence of fermion zero modes in various gravitational and membrane backgrounds is better understood  \cite{liu,oikonomou1}. As claimed and exemplified in \cite{rossi}, the existence of fermionic zero modes in the aforementioned backgrounds is always associated with some symmetry and in some particular cases this symmetry was a kind of supersymmetry (see \cite{rossi} for more details). This curious supersymmetry is not a symmetry of the Lagrangian itself but a symmetry of some field configurations satisfying some Ansatz. The appearance of fermionic zero modes is a direct consequence of this hidden symmetry. Along this line of research, we demonstrate that the fermionic zero modes are directly related to some $N=2$ supersymmetric quantum mechanical algebras which combine under certain circumstances to an extended SUSY QM structure. This can be potentially interesting, towards answering the question how zero modes occur in vortices and solitonic backgrounds. Our work may not provide a complete answer on this question, but provides useful insights on the Hilbert space structure of the fermionic zero modes and also for the Hilbert space geometry of these fermionic configurations. This in turn sheds some light on the full problem of the origin of fermionic zero modes in vortices backgrounds.

This paper is organized as follows: In section 1 we present the general theoretical framework of the fermionic system around the non-abelian vortices. In addition, we demonstrate how the two $N=2$ SUSY QM algebras are constructed and how the zero modes of the fermions constitute the Hilbert spaces of the supersymmetric quantum mechanical spaces. Moreover, we show how these algebras can be combined to form higher dimensional representations of an $N=2$ SUSY QM algebra. In section 2, we show how the two $N=2$, $d=1$ algebras can constitute an $N=4$, $d=1$ SUSY QM algebra with central charge. Finally, in section 3 we show that the Hilbert space of the fermionic zero modes is equipped with an $R$-symmetry which is the product of global $U(1)$ symmetries. The conclusions follow in the end of the article.

\section{Fermionic Zero Modes and Higher Order Representations of $N=2$ SUSY QM}

\subsection{Preliminaries}

Consider an $SU(2)$ gauge field coupled to two Higgs scalar fields $\phi$ and $\chi$, with the latter two in the adjoint representation of the aforementioned gauge group. The Lagrangian of the fermion-gauge-Higgs system is given by:
\begin{align}\label{lag122}
&\mathcal{L}=-\frac{1}{4}F_{\mu \nu}^aF^{\mu \nu}_a+|D_{\mu}\phi|^2+|D_{\mu}\chi|^2+c_2\phi^2-c_4\phi^4+d_2\chi^2-d_4\chi^4+e_2(\phi \cdot \chi)
\\ \notag & +\bar{\psi}i\gamma^{\mu}\partial_{\mu}\psi+g_1\bar{\psi}\phi_aT^a\psi+g_2\bar{\psi}\chi_aT^{a}\psi
\end{align}
Minimizing the potential, 
\begin{equation}\label{pot}
v(\chi,\phi)=c_2\phi^2-c_4\phi^4+d_2\chi^2-d_4\chi^4+e_2(\phi \cdot \chi)
\end{equation}
When $c_2d_2/c_4d_4>e_2^2/e_4^2$, the symmetry is broken down to $Z_2$ \cite{das}, and since the first homotopy group of the $Z_2$ quotient group of $SU(2)$ is $Z_2$, that is:
\begin{equation}\label{quotientsu2}
\pi_1(SU(2)/Z_2)=Z_2
\end{equation}
the physical system has $Z_2$-type topologically stable vortices. The minimal energy vortex solution is:
\begin{align}\label{jdhhgd}
& A_i^a=\delta_{i\theta}\delta_{a3}\frac{F(\rho-1)}{\rho},{\,}{\,}{\,}A_0=0\\ \notag &
\chi (x)=g(\rho )(\sin (n\theta),\cos (n\theta),0) \\ \notag &
\phi (x)=f(\rho) (\cos (n\theta),\sin (n\theta),0)
\end{align}
where $F(\rho )$ a real function with boundary values $F(0)=1$ and $F (\infty )=-n+1$. The number $n$ is the winding number and is related with the topologically conserved flux $\Phi =2\pi n (\mathrm{mod}2)$, $n=1,2,...\infty$. The Yukawa terms appearing in the Lagrangian (\ref{lag122}) render the fermions massive when the gauge symmetry is broken. Owing to the cylindrical symmetry of the gauge field and Higgs field configurations, the Dirac equation has fermionic solutions in the fundamental representation of the gauge group, of type (regarding spatial configurations only)  \cite{das}:
\begin{equation}\label{totalsol}
\psi (\rho,\theta,z)=e^{ikz}\left(%
\begin{array}{c}
  \psi_1 (\rho,\theta)\\
  \psi_2 (\rho,\theta) \\
\end{array}%
\right)
\end{equation}
Since the focus is on the zero modes, we may set $k=0$ without loosing generality \cite{das}. In equation (\ref{totalsol}), the $\psi_i(\rho,\theta)$ spinors are of the form:
\begin{align}\label{analsolution}
\psi_i(\rho,\theta)=\sum_{m=-\infty}^{\infty}a_m\psi_i^{(m)}(\rho,\theta)
\end{align}
The $\psi_i^{(m)}(\rho,\theta)$ spinors are equal to:
\begin{align}\label{analsolution}
&\psi_1^{(m)}(\rho,\theta)=\frac{1}{\sqrt{\rho}}\left(%
\begin{array}{c}
  e^{im\theta}\eta_1^{(m)}(\rho)\\
  e^{i(m+1)\theta}\eta_2^{(m)}(\rho) \\
\end{array}%
\right), \\ \notag &
\psi_2^{(m)}(\rho,\theta)=\frac{1}{\sqrt{\rho}}\left(%
\begin{array}{c}
  e^{i(m+n)\theta}\xi_1(\rho)\\
  e^{i(m+n+1)\theta}\xi_2(\rho) \\
\end{array}%
\right)
\end{align}
Having these in mind, the Dirac equation (when we consider zero modes) reduces to the following radial equations \cite{das}:
\begin{align}\label{1stsetofradeqns}
& \Big{[}\frac{\mathrm{d}}{\mathrm{d}\rho}+\frac{1}{\rho}(m+1-\frac{1}{2}F(\rho))\Big{]}\eta_2(\rho)=-X(\rho)\xi_1(\rho), \\ \notag &
\Big{[}\frac{\mathrm{d}}{\mathrm{d}\rho}-\frac{1}{\rho}(m+1-\frac{1}{2}F(\rho))\Big{]}\eta_1(\rho)=X(\rho)\xi_2(\rho), \\ \notag &
\Big{[}\frac{\mathrm{d}}{\mathrm{d}\rho}+\frac{1}{\rho}(m+n+\frac{1}{2}F(\rho))\Big{]}\eta_2(\rho)=X^*(\rho)\eta_1(\rho), \\ \notag &
\Big{[}\frac{\mathrm{d}}{\mathrm{d}\rho}-\frac{1}{\rho}(m+n+\frac{1}{2}F(\rho))\Big{]}\xi_1(\rho)=-X^*(\rho)\eta_2(\rho)
\end{align}
with $X(\rho)$ in general some complex arbitrary function of $\rho$ (as we shall see, the case in which $X(\rho)$ is real is particularly interesting). We shall take into account only normalizable solutions, a fact that is crucial for the analysis, since for these solutions, the operators we shall introduce later on this section are rendered Fredholm. As analyzed in \cite{das}, there are $n$ different normalizable zero modes solutions, with $n$ the topological winding number.

\subsection{$N=2$ SUSY QM in the Fermionic Sector}

Consider only a pair of the aforementioned well behaved normalizable zero modes, namely the ones corresponding to $m=-n+a$ and $m=-1+a$, with $a\geq 0$, for some value of the topological number $n$. These are explicitly of the form:
\begin{align}\label{analsolutiodhfhgfn}
&\psi_1^{a_1}(\rho,\theta)=\left(%
\begin{array}{c}
  0\\
  e^{i(-n+a+1)\theta}\eta_2^{a_1}(\rho) \\
\end{array}%
\right),{\,}{\,}{\,} \psi_2^{a_1}(\rho,\theta)=\left(%
\begin{array}{c}
  e^{ia\theta}\xi_1^{a_1}(\rho)\\
  0 \\
\end{array}%
\right){\,}{\,}{\,}(m=-n+a)\\ \notag &
\psi_1^{a_2}(\rho,\theta)=\left(%
\begin{array}{c}
  0\\
  e^{-ia\theta}\eta_2^{a_2}(\rho) \\
\end{array}%
\right),{\,}{\,}{\,} \psi_2^{a_2}(\rho,\theta)=\left(%
\begin{array}{c}
  e^{-i(-n+a+1)\theta}\xi_1^{a_2}(\rho)\\
  0 \\
\end{array}%
\right){\,}{\,}{\,}(m=-1-a)
\end{align}
Now the radial Dirac equations of motion takes the form:
\begin{align}\label{1stsetofradeqnsfdfd}
& \Big{[}\frac{\mathrm{d}}{\mathrm{d}\rho}+\frac{1}{\rho}(-n+a+1-\frac{1}{2}F(\rho))\Big{]}\eta_2^{a_1}(\rho)=-X(\rho)\xi_1^{a_1}(\rho), \\ \notag &
\Big{[}\frac{\mathrm{d}}{\mathrm{d}\rho}-\frac{1}{\rho}(a+\frac{1}{2}F(\rho))\Big{]}\xi_1^{a_1}(\rho)=-X^*(\rho)\eta_2^{a_1}(\rho), \\ \notag &
\Big{[}\frac{\mathrm{d}}{\mathrm{d}\rho}-\frac{1}{\rho}(a+\frac{1}{2}F(\rho))\Big{]}\eta_2^{a_2}(\rho)=-X(\rho)\xi_1^{a_2}(\rho), \\ \notag &
\Big{[}\frac{\mathrm{d}}{\mathrm{d}\rho}+\frac{1}{\rho}(-n+a+1-\frac{1}{2}F(\rho))\Big{]}\xi_1^{a_2}(\rho)=-X^*(\rho)\eta_2^{a_2}(\rho)
\end{align}
The equations (\ref{1stsetofradeqnsfdfd}) will be the starting point of our analysis. We focus on the first two equations which give the zero modes of the fields $\eta_2^{a_1}(\rho)$ and $\xi_1^{a_1}(\rho)$. As is established in \cite{das}, there exist exactly one pair of normalized zero modes corresponding to the fields $\eta_2^{a_1}(\rho)$ and $\xi_1^{a_1}(\rho)$. The normalization condition is necessary for our analysis since, as we shall see, the operators we shall encounter are Fredholm. Recall that these modes correspond to a certain number $m$, and there exist a total number of $n$ different zero modes (which means that we have an infinite tower of discrete modes). So what we discuss now for the fields $\eta_2^{a_1}(\rho)$ and $\xi_1^{a_1}(\rho)$, which correspond to a certain number $m$, hold true for all the $n$ different pairs of zero modes. From the first two equations of (\ref{1stsetofradeqnsfdfd}) we can form the operator $\mathcal{D}_{a_1}$,
\begin{equation}\label{susyqmrn567m}
\mathcal{D}_{a_1}=\left(%
\begin{array}{cc}
 \frac{\mathrm{d}}{\mathrm{d}\rho}+\frac{1}{\rho€}(-n+a+1-\frac{1}{2}F(\rho)) & -X(\rho)
 \\ -X^*(\rho) & \frac{\mathrm{d}}{\mathrm{d}\rho}-\frac{1}{\rho}(a+\frac{1}{2}F(\rho)) \\
\end{array}%
\right)
\end{equation}
which is considered to act on the vector:
\begin{equation}\label{ait34e1}
|\Psi_{a_1}\rangle =\left(%
\begin{array}{c}
  \eta_2^{a_1}(\rho) \\
   \xi_1^{a_1}(\rho) \\
\end{array}%
\right).
\end{equation}
Hence, the first two equations of (\ref{1stsetofradeqnsfdfd}) can be rewritten as follows:
\begin{equation}\label{transf}
\mathcal{D}_{a_1}|\Psi_{a_1}\rangle=0
\end{equation}
The solutions corresponding to the above relation (\ref{transf}) yield the zero modes of the first two equations of (\ref{1stsetofradeqnsfdfd}). As we already mentioned, there is only one pair of zero modes corresponding to the functions $\eta_2^{a_1}(\rho)$ and $\xi_1^{a_1}(\rho)$, for a ceratin $m$, hence we may easily conclude that:
\begin{equation}\label{dimeker}
\mathrm{dim}{\,}\mathrm{ker}\mathcal{D}_{a_1}=1
\end{equation}
The adjoint of the operator $\mathcal{D}_{a_1}$, is equal to:
\begin{equation}\label{eqndag}
\mathcal{D}_{a_1}^{\dag}=\left(%
\begin{array}{cc}
 \frac{\mathrm{d}}{\mathrm{d}\rho}+\frac{1}{\rho€}(-n+a+1-\frac{1}{2}F(\rho)) & -X(\rho)
 \\ -X^*(\rho) & \frac{\mathrm{d}}{\mathrm{d}\rho}-\frac{1}{\rho}(a+\frac{1}{2}F(\rho)) \\
\end{array}%
\right)
\end{equation}
Obviously, the kernel of the adjoint operator $\mathcal{D}_{a_1}^{\dag}$, is equal to:
\begin{equation}\label{dimeke1r11}
\mathrm{dim}{\,}\mathrm{ker}\mathcal{D}_{a_1}^{\dag}=1
\end{equation}
The fermionic zero modes around vortices $\eta_2^{a_1}(\rho)$ and $\xi_1^{a_1}(\rho)$, constitute an unbroken $N=2$, $d=1$ SUSY QM algebra. The supercharges of this one dimensional $N=2$, SUSY algebra are related to the operator $\mathcal{D}_{A_1}$ in the following way:
\begin{equation}\label{s7}
\mathcal{Q}_{a_1}=\bigg{(}\begin{array}{ccc}
  0 & \mathcal{D}_{a_1} \\
  0 & 0  \\
\end{array}\bigg{)},{\,}{\,}{\,}\mathcal{Q}^{\dag}_{a_1}=\bigg{(}\begin{array}{ccc}
  0 & 0 \\
  \mathcal{D}_{a_1}^{\dag} & 0  \\
\end{array}\bigg{)}
\end{equation}
and the corresponding quantum Hamiltonian of the system is:
\begin{equation}\label{s11}
\mathcal{H}_{a_1}=\bigg{(}\begin{array}{ccc}
 \mathcal{D}_{a_1}\mathcal{D}_{a_1}^{\dag} & 0 \\
  0 & \mathcal{D}_{a_1}^{\dag}\mathcal{D}_{a_1}  \\
\end{array}\bigg{)}
\end{equation}
The aforementioned operators, namely $\mathcal{Q}_{a_1}$, $\mathcal{Q}^{\dag}_{a_1}$ and $\mathcal{H}_{a_1}$, satisfy the one dimensional SUSY QM algebra:
\begin{equation}\label{relationsforsusy}
\{\mathcal{Q}_{a_1},\mathcal{Q}^{\dag}_{a_1}\}=\mathcal{H}_{a_1}{\,}{\,},\mathcal{Q}_{a_1}^2=0,{\,}{\,}{\mathcal{Q}_{a_1}^{\dag}}^2=0
\end{equation}
The total Hilbert space of the SUSY QM system, which we denote $\mathcal{H}$, is $Z_2$-graded by the operator $\mathcal{W}$, known as ''Witten parity''. This operator has the following commutation and anti-commutation properties,
\begin{equation}\label{s45}
[\mathcal{W},\mathcal{H}_{a_1}]=0,{\,}{\,}{\,}\{\mathcal{W},\mathcal{Q}_{a_1}\}=\{\mathcal{W},\mathcal{Q}_{a_1}^{\dag}\}=0
\end{equation}
and satisfies the following identity,
\begin{equation}\label{s6}
\mathcal{W}^{2}=1
\end{equation}
In addition, in our case it can be represented by the following matrix:
\begin{equation}\label{wittndrf}
\mathcal{W}=\bigg{(}\begin{array}{ccc}
  1 & 0 \\
  0 & -1  \\
\end{array}\bigg{)}
\end{equation}
With the $Z_2$-grading the involution operator $W$ provides, the total Hilbert space $\mathcal{H}$, of the SUSY QM system can be written as:
\begin{equation}\label{fgjhil}
\mathcal{H}=\mathcal{H}^+\oplus \mathcal{H}^-
\end{equation}
The vectors belonging to the two subspaces $\mathcal{H}^{\pm}$, are classified to Witten parity even and Witten parity odd parity states:
\begin{equation}\label{shoes}
\mathcal{H}^{\pm}=\mathcal{P}^{\pm}\mathcal{H}=\{|\psi\rangle :
\mathcal{W}|\psi\rangle=\pm |\psi\rangle \}
\end{equation}
Moreover, the Hamiltonians of the graded Hilbert spaces are:
\begin{equation}\label{h1}
{\mathcal{H}}_{+}=\mathcal{D}_{a_1}{\,}\mathcal{D}_{a_1}^{\dag},{\,}{\,}{\,}{\,}{\,}{\,}{\,}{\mathcal{H}}_{-}=\mathcal{D}_{a_1}^{\dag}{\,}\mathcal{D}_{a_1}
\end{equation}
Using the representation (\ref{wittndrf}) for the Witten parity operator,
the Witten parity eigenstates can represented by the vectors,
\begin{equation}\label{phi5}
|\psi^{+}\rangle =\left(%
\begin{array}{c}
  |\phi^{+}\rangle \\
  0 \\
\end{array}%
\right),{\,}{\,}{\,}
|\psi^{-}\rangle =\left(%
\begin{array}{c}
  0 \\
  |\phi^{-}\rangle \\
\end{array}%
\right)
\end{equation}
with $|\phi^{\pm}\rangle$ $\epsilon$ $\mathcal{H}^{\pm}$. It is an easy task to write the fermionic states of the fermion vortex system in terms of the SUSY QM algebra. So we have:
\begin{equation}\label{fdgdfgh}
|\Psi_{a_1}\rangle =|\phi^{-}\rangle=\left(%
\begin{array}{c}
   \eta_2^{a_1}(\rho) \\
   \xi_1^{a_1}(\rho) \\
\end{array}%
\right),{\,}{\,}{\,}|\Psi_{a_1}'\rangle =|\phi^{+}\rangle=\left(%
\begin{array}{c}
   \eta_2^{a_1}(\rho) \\
   \xi_1^{a_1}(\rho) \\
\end{array}%
\right)
\end{equation}
Hence, the associated even and odd parity SUSY QM states are:
\begin{equation}\label{phi5}
|\psi^{+}\rangle =\left(%
\begin{array}{c}
  |\Psi_{a_1}'\rangle \\
  0 \\
\end{array}%
\right),{\,}{\,}{\,}
|\psi^{-}\rangle =\left(%
\begin{array}{c}
  0 \\
  |\Psi_{a_1}\rangle \\
\end{array}%
\right)
\end{equation}
Supersymmetry is considered to be unbroken if the Witten index $\Delta$, is a non-zero integer, with $\Delta$,
\begin{equation}\label{phil}
\Delta =n_{-}-n_{+}
\end{equation}
when Fredholm operators are taken into account. In relation (\ref{phil}), $n_{\pm}$ denote the number of zero
modes of ${\mathcal{H}}_{\pm}$ in the subspace $\mathcal{H}^{\pm}$, with the constraint that these are finitely many. The operators we consider satisfy the latter constraint, since they have finite kernels, as can be seen from relations (\ref{dimeker}) and (\ref{dimeke1r11}).
Supersymmetry can also be unbroken in the case the Witten index is zero. This happens when $n_{+}=
n_{-}\neq 0$, which is actually what happens in our case. Supersymmetry is broken when $n_{+}=n_{-}=0$. 
The Witten index is directly connected to the Fredholm index of the operator $\mathcal{D}_{a_1}$ (recall that the Fredholm property is guaranteed from the finiteness of the kernels), as can be seen in the following equations:
\begin{align}\label{ker1}
&\Delta=\mathrm{dim}{\,}\mathrm{ker}
{\mathcal{H}}_{-}-\mathrm{dim}{\,}\mathrm{ker} {\mathcal{H}}_{+}=
\mathrm{dim}{\,}\mathrm{ker}\mathcal{D}_{LG}^{\dag}\mathcal{D}_{LG}-\mathrm{dim}{\,}\mathrm{ker}\mathcal{D}_{LG}\mathcal{D}_{LG}^{\dag}=
\\ \notag & \mathrm{ind} \mathcal{D}_{LG} = \mathrm{dim}{\,}\mathrm{ker}
\mathcal{D}_{LG}-\mathrm{dim}{\,}\mathrm{ker} \mathcal{D}_{LG}^{\dag}
\end{align}
Recalling the results of equations (\ref{dimeker}) and (\ref{dimeke1r11}), the Witten index is equal to:
\begin{equation}\label{fnwitten}
\Delta =0
\end{equation}
with $n_{+}=
n_{-}\neq 0$. Therefore, the fermionic system $\eta_2^{a_1}(\rho)$ and $\xi_1^{a_1}(\rho)$, in the presence of non-abelian vortices, has an unbroken $N=2$, $d=1$ supersymmetry. Hence, although the initial system has no connection with global spacetime supersymmetry, the fermionic zero modes have an unbroken supersymmetry. This fact however should not surprise us too much, since global spacetime supersymmetry and SUSY QM are in general two different concepts. Particularly, the SUSY QM supercharges
do not generate in any way spacetime supersymmetric transformations. In addition, in
SUSY QM, fermions and bosons are not directly related and are not classified as representations of the super-Poincare algebra in four dimensions. Moreover, the SUSY QM supercharges do not generate transformations between fermions and bosons but provide a $Z_2$-grading of the quantum Hilbert space of quantum states. Although global spacetime supersymmetry and SUSY QM are different concepts, there is some connection which was point out in the literature \cite{ivanov}. Particularly, extended (with $N = 4, 6...$) supersymmetric quantum mechanics models can describe
the dimensional reduction to one (temporal) dimension of N = 2 and N = 1
Super-Yang Mills models. This fact is an indication that extended SUSY QM could be a remnant of higher global spacetime supersymmetry, but there is no indication towards the connection of $N=2$ SUSY QM to some higher global supersymmetric algebra.

Similar considerations may lead to the conclusion that the second set of fermionic fields $\eta_2^{a_2}(\rho)$ and $\xi_1^{a_2}(\rho)$ are related to an unbroken $N=2$ SUSY QM algebra. Particularly, the last two equations of relation (\ref{1stsetofradeqnsfdfd}) can be written in terms of the operator $\mathcal{D}_{a_a}$,
\begin{equation}\label{susyqmrn567m}
\mathcal{D}_{a_2}=\left(%
\begin{array}{cc}
 \frac{\mathrm{d}}{\mathrm{d}\rho}+\frac{1}{\rho}(-n+a+1-\frac{1}{2}F(\rho)) & -X^*(\rho)
 \\ -X(\rho) & \frac{\mathrm{d}}{\mathrm{d}\rho}-\frac{1}{\rho}(a+\frac{1}{2}F(\rho)) \\
\end{array}%
\right)
\end{equation}
which is considered to act on the vector:
\begin{equation}\label{ait34e1}
|\Psi_{a_a}\rangle =\left(%
\begin{array}{c}
  \xi_1^{a_2}(\rho) \\
   \eta_2^{a_2}(\rho) \\
\end{array}%
\right).
\end{equation}
Therefore, the last two equations of (\ref{1stsetofradeqnsfdfd}) can be written as follows:
\begin{equation}\label{transf}
\mathcal{D}_{a_2}|\Psi_{a_2}\rangle=0
\end{equation}
In this case too, the kernels of the operators $\mathcal{D}_{a_2}$ and $\mathcal{D}^{\dag}_{a_2}$ are:
\begin{equation}\label{dimekerugigy}
\mathrm{dim}{\,}\mathrm{ker}\mathcal{D}_{a_2}=1,{\,}{\,}{\,}\mathrm{dim}{\,}\mathrm{ker}\mathcal{D}^{\dag}_{a_2}=1
\end{equation}        
The $N=2$ SUSY QM algebra is unbroken in this case for the same reasons as in the previous case. The essentials of the algebra are given by the following operators, namely the supercharges: 
\begin{equation}\label{s7}
\mathcal{Q}_{a_2}=\bigg{(}\begin{array}{ccc}
  0 & \mathcal{D}_{a_2} \\
  0 & 0  \\
\end{array}\bigg{)},{\,}{\,}{\,}\mathcal{Q}^{\dag}_{a_2}=\bigg{(}\begin{array}{ccc}
  0 & 0 \\
  \mathcal{D}_{a_2}^{\dag} & 0  \\
\end{array}\bigg{)}
\end{equation}
and the quantum Hamiltonian:
\begin{equation}\label{s11}
\mathcal{H}_{a_2}=\bigg{(}\begin{array}{ccc}
 \mathcal{D}_{a_2}\mathcal{D}_{a_2}^{\dag} & 0 \\
  0 & \mathcal{D}_{a_2}^{\dag}\mathcal{D}_{a_2}  \\
\end{array}\bigg{)}
\end{equation}
which satisfy the one dimensional SUSY QM algebra:
\begin{equation}\label{relationsforsusy}
\{\mathcal{Q}_{a_2},\mathcal{Q}^{\dag}_{a_2}\}=\mathcal{H}_{a_2}{\,}{\,},\mathcal{Q}_{a_2}^2=0,{\,}{\,}{\mathcal{Q}_{a_2}^{\dag}}^2=0.
\end{equation}
In conclusion, since the second set of fermionic fields also constitutes an unbroken $N=2$ SUSY QM algebra a question rises whether these two unbroken $N=2$, $d=1$ supersymmetries combine to form a higher order extended supersymmetry. The answer lies to the affirmative as we shall evince in a later section. Before getting into this, we shall investigate what are the characteristics of higher order representations of $N=2$ SUSY QM, formed by these two $N=2$, $d=1$ supersymmetries we studied in this section. This is the topic of the next section.

As a last comment before we close this section, note that the two supersymmetries we found correspond to the topological quantum number $m$ (related to the winding number $n$) of the initial fermion-non-abelian vortex system. We denote with $\mathcal{N}_m$ the total supersymmetry of the $m$-th level. As we established, this total supersymmetry is composed by two supersymmetries as follows:
\begin{equation}\label{sssuy}
\mathcal{N}_m=\mathcal{N}^{a_1}_m\oplus \mathcal{N}^{a_2}_m
\end{equation}
with $\mathcal{N}^{a_1}_m$ corresponding to the set $\eta_2^{a_1}(\rho)$ and $\xi_1^{a_1}(\rho)$ and $\mathcal{N}^{a_2}_m$ related to the fields $\eta_2^{a_2}(\rho)$ and $\xi_1^{a_2}(\rho)$. Therefore, since each level has two such supersymmetries, the total system has $n$ different supersymmetries. Denoting with $\mathcal{N}_{tot}$ the total supersymmetry, we have:
\begin{equation}\label{totalsusye}
\mathcal{N}_{tot}=\sum_{i=1}^{n}\mathcal{N}_i
\end{equation}
Thereby, we should have in mind that all the results holding true for the $m$-th level fermion system, hold true for all the other $(n-1)$-fermionic subsystems. Since $n$ takes values $n=1,2,...\infty$, we have an infinite set of $N=2$, $d=1$ supersymmetric subsystems.

\subsection{Higher Reducible Representations of $N=2$ SUSY QM}

\noindent We consider the supercharges $(\mathcal{Q}_{a_1},\mathcal{Q}_{a_2})$ and the corresponding operators $({\mathcal{D}}_{a_1},{\mathcal{D}}_{a_2})$. These supercharges and operators correspond to some number $m$ of zero modes and recall that we have $n$ such tuples of such operators. Every result holding true for the operators under study, hold true for every set of operators corresponding to some other number $n$. The two $N=2$ supersymmetries corresponding to the supercharges $(\mathcal{Q}_{a_1},\mathcal{Q}_{a_2})$, can be combined to a higher reducible representation of a single $N=2$, $d=1$ supersymmetry. Denoting the supercharges of the higher order representation by ${\mathcal{Q}}_{h}$ and  ${\mathcal{Q}}_{h}^{\dag}$, we can form them in the following manner:
\begin{equation}\label{connectirtyrtons}
{\mathcal{Q}}_{h}= \left ( \begin{array}{cccc}
  0 & 0 & 0 & 0 \\
  {\mathcal{D}}_{a_2} & 0 & 0 & 0 \\
0 & 0 & 0 & 0 \\
0 & 0 & {\mathcal{D}}_{a_1}^{\dag} & 0  \\
\end{array} \right),{\,}{\,}{\,}{\,}{\mathcal{Q}}_{h}^{\dag}= \left ( \begin{array}{cccc}
  0 &  {\mathcal{D}}_{a_2}^{\dag} & 0 & 0 \\
  0 & 0 & 0 & 0 \\
0 & 0 & 0 & {\mathcal{D}}_{a_1} \\
0 & 0 & 0 & 0  \\
\end{array} \right)
.\end{equation}
Moreover, the Hamiltonian of the higher order quantum system, which we denote $H_h$, reads,
\begin{equation}\label{connections1dtr}
H_{h}= \left ( \begin{array}{cccc}
  {\mathcal{D}}_{a_2}^{\dag}{\mathcal{D}}_{a_2} & 0 & 0 & 0 \\
  0 & {\mathcal{D}}_{a_2}{\mathcal{D}}_{a_2}^{\dag} & 0 & 0 \\
0 & 0 & {\mathcal{D}}_{a_1}{\mathcal{D}}_{a_1}^{\dag} & 0 \\
0 & 0 & 0 & {\mathcal{D}}_{a_1}^{\dag}{\mathcal{D}}_{a_1}  \\
\end{array} \right)
.\end{equation}
The operators of relations (\ref{connectirtyrtons}) and (\ref{connections1dtr}), satisfy the $N=2$, $d=1$ SUSY QM algebra, that is:
\begin{equation}\label{mousikisimagne}
\{ {\mathcal{Q}}_{h},{\mathcal{Q}}_{h}^{\dag}\}=H_{h},{\,}{\,}{\mathcal{Q}}_{h}^2=0,{\,}{\,}{{\mathcal{Q}}_{h}^{\dag}}^2=0,{\,}{\,}\{{\mathcal{Q}}_{h},\mathcal{W}_{h}\}=0,{\,}{\,}\mathcal{W}_{h}^2=I,{\,}{\,}[\mathcal{W}_{h},H_{h}]=0
.\end{equation}
The operator $W_h$, appearing above, is the generalized Witten parity operator which in this case is equal to:
\begin{equation}\label{wparityopera}
\mathcal{W}_{h}= \left ( \begin{array}{cccc}
  1 & 0 & 0 & 0 \\
  0 & -1 & 0 & 0 \\
0 & 0 & 1 & 0 \\
0 & 0 & 0 & -1  \\
\end{array} \right)
.\end{equation}
Equivalent representations can be formed by making the following substitutions:
\begin{equation}\label{setof transformations}
\mathrm{Set}{\,}{\,}{\,}A:{\,}
\begin{array}{c}
 {\mathcal{D}}_{a_2}\rightarrow {\mathcal{D}}_{a_2}^{\dag} \\
  {\mathcal{D}}_{a_1}^{\dag}\rightarrow {\mathcal{D}}_{a_1} \\
\end{array},{\,}{\,}{\,}\mathrm{Set}{\,}{\,}{\,}B:{\,}
\begin{array}{c}
 {\mathcal{D}}_{a_2}\rightarrow {\mathcal{D}}_{a_1}^{\dag} \\
  {\mathcal{D}}_{a_1}^{\dag}\rightarrow {\mathcal{D}}_{a_2} \\
\end{array},{\,}{\,}{\,}\mathrm{Set}{\,}{\,}{\,}C:{\,}
\begin{array}{c}
 {\mathcal{D}}_{a_2}\rightarrow {\mathcal{D}}_{a_1} \\
  {\mathcal{D}}_{a_1}^{\dag}\rightarrow {\mathcal{D}}_{a_2}^{\dag} \\
\end{array}
.\end{equation}
In addition, another higher order reducible representation of the $N=2$ SUSY QM algebra, equivalent to (\ref{connectirtyrtons}), is given by: 
\begin{equation}\label{connectirtyfhfghrtons}
{\mathcal{Q}}_{h}= \left ( \begin{array}{cccc}
  0 & 0 & 0 & 0 \\
  0 & 0 & 0 & 0 \\
{\mathcal{D}}_{a_2} & 0 & 0 & 0 \\
0 & {\mathcal{D}}_{a_1}^{\dag} & 0 & 0  \\
\end{array} \right),{\,}{\,}{\,}{\,}{\mathcal{Q}}_{h}^{\dag}= \left ( \begin{array}{cccc}
  0 & 0 & {\mathcal{D}}_{a_2}^{\dag} & 0 \\
   & 0 & 0 & {\mathcal{D}}_{a_1} \\
0 & 0 & 0 & 0 \\
0 & 0 & 0 & 0  \\
\end{array} \right)
.\end{equation}
The corresponding Hamiltonian is equal to,
\begin{equation}\label{connectihgghdhtons1dtr}
H_{h}= \left ( \begin{array}{cccc}
  {\mathcal{D}}_{a_2}^{\dag}{\mathcal{D}}_{a_2} & 0 & 0 & 0 \\
  0 & {\mathcal{D}}_{a_1}^{\dag}{\mathcal{D}}_{a_1} & 0 & 0 \\
0 & 0 & {\mathcal{D}}_{a_1}{\mathcal{D}}_{a_1}^{\dag} & 0 \\
0 & 0 & 0 & {\mathcal{D}}_{a_2}{\mathcal{D}}_{a_2}^{\dag}  \\
\end{array} \right)
.\end{equation}
In conclusion we have $n$ different higher order representations of $N=2$, $d=1$ algebras, formed by operators of the form (\ref{connectirtyrtons}) and (\ref{connections1dtr}). Apart from these reducible higher order representations of $N=2$, $d=1$ supersymmetry we just presented, a question than naturally springs to mind is whether this fermionic system around the non-Abelian vortex is in some way related to any extended supersymmetry with $N>2$. This is because there exist two $N=2$, $d=1$ supersymmetries with zero central charge, which in some way can be combined to give an enhanced supersymmetric structure. The answer to the question lies to the affirmative and this is the subject of the next section.

\section{Central Charge Extended $N=4$ SUSY QM for all Fermion Pairs}

Motivated by the existence of two $N=2$ SUSY QM algebras underlying the system of fundamental fermions around non-Abelian vortices, in this section we shall search if there is a more rich supersymmetry structure underlying the fermion system. In fact, as we shall demonstrate, the system has an central charge extended $N=4$ supersymmetry. This result occurs in the case that $X(\rho)$ is a real function of $\rho$ with asymptotic behavior that yields normalizable zero modes for the fermions \cite{das}. In order to see this, we compute the following commutation and anti-commutation relations:
\begin{align}\label{commutatorsanticomm}
&\{{{\mathcal{Q}}_{a_1}},{{\mathcal{Q}}^{\dag}_{a_1}}\}=2\mathcal{H},{\,}\{{{\mathcal{Q}}_{a_2}},{{\mathcal{Q}}^{\dag}_{a_2}}\}=2\mathcal{H},{\,}\{{{\mathcal{Q}}_{a_2}},{{\mathcal{Q}}_{a_2}}\}=0,{\,}\{{{\mathcal{Q}}_{a_1}},{{\mathcal{Q}}_{a_1}}\}=0,{\,}{\,}\\
\notag & \{{{\mathcal{Q}}_{a_2}},{{\mathcal{Q}}^{\dag}_{a_1}}\}=\mathcal{Z},{\,}\{{{\mathcal{Q}}_{a_1}},{{\mathcal{Q}}^{\dag}_{a_2}}\}=\mathcal{Z},{\,}\\ \notag
&\{{{\mathcal{Q}}^{\dag}_{a_1}},{{\mathcal{Q}}_{a_1}}^{\dag}\}=0,\{{{\mathcal{Q}}^{\dag}_{a_2}},{{\mathcal{Q}}^{\dag}_{a_2}}\}=0,{\,}\{{{\mathcal{Q}}^{\dag}_{a_2}},{{\mathcal{Q}}^{\dag}_{a_1}}\}=0,{\,}\{{{\mathcal{Q}}_{a_2}},{{\mathcal{Q}}_{a_1}}\}=0{\,}\\
\notag
&[{{\mathcal{Q}}_{a_1}},{{\mathcal{Q}}_{a_2}}]=0,[{{\mathcal{Q}}^{\dag}_{a_2}},{{\mathcal{Q}}^{\dag}_{a_1}}]=0,{\,}[{{\mathcal{Q}}_{a_1}},{{\mathcal{Q}}_{a_1}}]=0{\,}[{{\mathcal{Q}}^{\dag}_{a_1}},{{\mathcal{Q}}^{\dag}_{a_1}}]=0,{\,}\\
\notag &
[{\mathcal{H}}_{a_2},{{\mathcal{Q}}_{a_1}}]=0,{\,}[{\mathcal{H}}_{a_2},{{\mathcal{Q}}^{\dag}_{a_1}}]=0,{\,}[\mathcal{H}_{a_1},{{\mathcal{Q}}^{\dag}_{a_2}}]=0,{\,}[\mathcal{H}_{a_1},{{\mathcal{Q}}_{a_2}}]=0,{\,}
\end{align}
with $\mathcal{Z}$, the central charge of the above $N=4$ SUSY QM algebra, which is:
\begin{equation}\label{zcentralcharge}
\mathcal{Z}=2\mathcal{H}_{a_1}=2{\mathcal{H}}_{a_2}=2\mathcal{H}
\end{equation}
The central charge $\mathcal{Z}$ commutes with all the operators of the two $N=2$, $d=1$ SUSY QM algebras, that is, with the
supercharges ${{\mathcal{Q}}_{a_2}},{{\mathcal{Q}}_{a_1}}$, their conjugates ${{\mathcal{Q}}^{\dag}_{a_2}},{{\mathcal{Q}}^{\dag}_{a_1}}$ and the Hamiltonians, $\mathcal{H}=\mathcal{H}_{a_1}={\mathcal{H}}_{a_2}$.
The relations (\ref{commutatorsanticomm}) correspond to a central charge extended $N=4$ SUSY QM
algebra and particularly to an algebra with two central charges. Indeed, for an $N=4$ algebra we have:
\begin{align}\label{n4algbe}
&\{Q_i,Q_j^{\dag}\}=2\delta_i^jH+Z_{ij},{\,}{\,}i=1,2 \\ \notag &
\{Q_i,Q_j\}=0,{\,}{\,}\{Q_i^{\dag},Q_j^{\dag}\}=0
\end{align}
The algebra (\ref{commutatorsanticomm}) possesses two central charges which
are equal and specifically, these charges are equal to $Z_{12}=Z_{21}=\mathcal{Z}$.

\noindent The existence of an $N=4$ SUSY QM algebra underlying the fermionic system is particularly interesting since important information about the Hilbert space of the fermionic zero modes and possible connections with more involved theories  may be revealed. The $N=4$ supersymmetric algebra is important in string theory, since extended SUSY QM models with $N\geq 4$, result from the dimensional reduction to one dimension of $N=2$ and $N=1$ Super-Yang Mills theories \cite{ivanov}. In addition, the $N=4$ SUSY QM structure can be directly related to a generalized harmonic superspace, with the latter being a useful tool for $N\geq 4$ supersymmetric model building.

Before closing this section, recall that each level of fermion fields with a certain $m$ (see section 1), has an underlying $N=4$ SUSY QM structure, and therefore in the end we have infinite tuples of different $N=4$, $d=1$ supersymmetries.

\subsection{A Brief Discussion on Topology and the Existence of Extended SUSY QM}

As was shown in \cite{das}, when $n$ is even the zero modes of the fields $\psi_i^{a_j}$ appearing in relation (\ref{analsolutiodhfhgfn}), pair up, while for $n$ odd, there is always a zero mode which is not paired up \cite{das}. As pointed out in \cite{das}, this fact suggests in some way some connection with the topology of the background field since only vortices with $n$ odd are topologically stable. However this connection was not made rigid in terms of some index theorem. Perhaps our results could shed some light towards this direction. Particularly, the calculation of the Witten index gives us some topological information for each level of zero modes, which is characterized by the number $m$. In addition, the existence of two $N=2$, $d=1$ supersymmetries in the fermionic zero modes corresponding to the fields $\psi_i^{a_j}$ and specifically the fact that these two combine to form an extended supersymmetric structure cannot be accidental. This probably suggests that there is some underlying geometrical reason connected with the Hilbert space of the fermions in the vortices background, since the two Hilbert spaces corresponding to the $N=2$, $d=1$ supersymmetries combine to form an $N=4$ central charge extended one. Hence, the total Hilbert space is much more involved than it originally appears and the existence of SUSY QM might prove useful towards this line of research. This however is beyond the scope of this paper.

\section{Global $U(1)$ R-parities in the Fundamental Fermions Sector}

The existence of two $N=2$ SUSY QM algebras on each fermionic zero mode set corresponding to the quantum number $m$, has an implication on the Hilbert space of quantum states. Particularly, the Hilbert space quantum states, that are closely related to the fermionic zero modes wave functions, are invariant under a product of two global $U(1)$ symmetries. In order to see that, let us focus on the SUSY QM algebra with supercharges $\mathcal{Q}_{a_1}$ and $\mathcal{Q}^{\dag}_{a_1}$. We perform the following transformation on the supercharges $\mathcal{Q}_{a_1}$ and $\mathcal{Q}^{\dag}_{a_1}$:
\begin{align}\label{transformationu1}
& {\mathcal{Q}}_{a_1}^{'}=e^{-ia}{\mathcal{Q}}_{a_1}, {\,}{\,}{\,}{\,}{\,}{\,}{\,}{\,}
{\,}{\,}{\mathcal{Q}}^{'\dag}_{a_1}=e^{ia}{\mathcal{Q}}^{\dag}_{a_1}
.\end{align}
The Hamiltonian of the system $\mathcal{H}_{a_1}$ is invariant under this global $U(1)$ transformation, but the quantum states of the system are transformed according to their $Z_2$ grading. We denote the global $U(1)$ transformation corresponding to the supercharges under study, $U_{a_1}(1)$. Recall that the graded quantum states are the ones of relation (\ref{phi5}) $|\psi^{+}\rangle$ $\in$ $\mathcal{H}^{+}$ and
$|\psi^{-}\rangle$ $\in$ $\mathcal{H}^{-}$, that is, the Witten parity even and odd states respectively. Under the global $U_{a_1}(1)$, the quantum states are transformed as follows:
\begin{equation}
|\psi^{'+}\rangle=e^{-i\beta_{+}}|\psi^{+}\rangle,
{\,}{\,}{\,}{\,}{\,}{\,}{\,}{\,}
{\,}{\,}|\psi^{'-}\rangle=e^{-i\beta_{-}}|\psi^{-}\rangle
.\end{equation}
Of course, the parameters $\beta_{+}$ and $\beta_{-}$ are global
parameters that satisfy the relation $a=\beta_{+}-\beta_{-}$. Hence, having in mind that there are two more supercharges $\mathcal{Q}_{a_2}$ and $\mathcal{Q}_{a_2}^{\dag}$, the total global symmetry of the system is $U_{a_1}(1)\times U_{a_2}(1)$. Note that taking into account the fact that we have $n$ different zero modes, the total global symmetry of the system, which we denote $\mathcal{R}_{tot}$, is of the form:
\begin{equation}\label{tottglob}
\mathcal{R}_{tot}=\underbrace{U_{a_1}(1)\times U_{a_2}\times...\times U_{a_{n1}}(1)\times U_{a_{n2}}(1)}_{n{\,}\mathrm{times}}
\end{equation}

\section*{Concluding Remarks}

In this article we demonstrated that the fermionic zero modes of the fermions in the fundamental representation (isospinor fermions) constitute two independent $N=2$, $d=1$ supersymmetries. These supersymmetries are combined, under certain circumstances, to give an enhanced extended supersymmetric structure and particularly an $N=4$, $d=1$ SUSY QM algebra with central charge. These findings can be useful when one considers the problem of finding the origin of fermionic zero modes around vortices and in general, around solitonic backgrounds. In principle, the complexity of a general problem may be drastically reduced if we find ways to recognize internal symmetries, or if a classification according to an already known symmetry is possible. In addition, if we find instead of symmetries, any regularity or repeating pattern, we gain even more deep insight to the problem at hand. As it was pointed out in reference \cite{rossi}, the existence of the fermionic zero modes can be associated to the existence of a hidden symmetry that the system has. This symmetry is a kind of supersymmetry. It is therefore very intriguing that the zero modes of the system are directly associated to $N=2$ and $N=4$ one dimensional supersymmetries. These SUSY QM algebras can be remnants of this hidden supersymmetry in some way. This study is rather interesting and we hope to address these issues in the future.

\section*{Acknowledgments} Financial support by the Research
Committee of the Technological Education Institute of Serres,
under grant SAT/ME/201113-23/05, is gratefully acknowledged.

\end{document}